\documentclass[12pt]{article}
\usepackage{graphicx}
\usepackage{cite}
\usepackage{pstricks}
\usepackage{amsfonts}
\begin{document}

\begin{titlepage}

\vskip 2pc plus 1fil minus 0.5pc

\centerline{\Large \bf Dark Matter in Minimal Trinification$\strut^{\ast}$}
\vskip 1pc minus 0.5pc
\centerline{\large E. Di Napoli$\strut^{\rm (1)}$, D. George$\strut^{\rm (2)}$, 
     M. Hertzberg$\strut^{\rm (3)}$, F. Metzler$\strut^{\rm (4)}$ 
     and E. Siegel$\strut^{\rm (5)}$}
\vskip 1pc minus 0.5pc
\tabskip=0pt
\setbox0=\vtop{\halign{%
    \small\it #\hfil\cr
    \noalign{\vskip 1pc}
    \llap{\rm (1)\enspace}\relax
    University of North Carolina,\cr
    Department of Physics and Astronomy,\cr
    CB 3255 Phillips Hall,\cr
    Chapel Hill NC 27599-3255, USA\cr
    {\tt edodin@physics.unc.edu}\cr
    \noalign{\vskip 0.5pc}
    \llap{\rm (3)\enspace}\relax
     MIT, Center for Theoretical Physics,\cr
    77, Massachusetts Av.,\cr
     Cambridge MA 02139, USA\cr
    {\tt mphertz@mit.edu}\cr
    \noalign{\vskip 0.5pc}
    \llap{\rm (5)\enspace}\relax
    University of Wisconsin,\cr
    Department of Physics,\cr
    1150 University Avenue,\cr
    Madison WI 53706, USA\cr
    {\tt siegel@phys.ufl.edu}\cr
    }}
\setbox1=\vtop{\halign{%
    \small\it #\hfil\cr
    \noalign{\vskip 1pc}
    \llap{\rm (2)\enspace}\relax
    The School of Physics,\cr
    University of Melbourne,\cr
    Victoria, 3010 Australia\cr
    {\tt d.george@physics.unimelb.edu.au}\cr
    \noalign{\vskip 0.5pc}
    \llap{\rm (4)\enspace}\relax
     Institut f\"ur Theoretische Teilchenphysik,\cr
    Universit\"at Karlsruhe,\cr
    76128 Karlsruhe, Germany\cr
    {\tt metzler@particle.uni-karlsruhe.de}\cr
    }}
\centerline{\box0 \hskip 4pc plus 1fill \box1}

\vskip 2pc plus 1fil minus 0.5pc

\centerline{ABSTRACT}
\smallskip
We study an example of  Grand Unified Theory (GUT), known as 
trinification, which was first introduced in 1984 by S.Glashow. 
This model has the GUT gauge group
as $[SU(3)]^3$ with a discrete $\mathbb{Z}_3$ to ensure the couplings 
are unified at the GUT scale. In this letter we consider this 
trinification model in its minimal formulation and investigate its
robustness in the context of cosmology. In particular we show that 
for a large set of the parameter space the model doesn't seem to
provide a Dark Matter candidate compatible with cosmological data. 
\par
\vskip 2pc plus 1fil minus 0.5pc

\begingroup
    \catcode`\@=11
    \let\@thefnmark=\relax
    \@footnotetext{%
        \nobreak
        \par\noindent \hangafter=1 \hangindent=\parindent
        {\large $\ast$}\enspace
        Article based on research supported by the US Department of
        Energy (grant DE--FG02--06ER41418)
        }
\endgroup

\end{titlepage}
\newpage
\pagenumbering{arabic}

\section{Introduction}
In this letter we would like to present the preliminary and
suggestive results of a more ambitious and extensive research
project. We study an example of Grand Unified
Theory (GUT) in the context of  certain requirements dictated by cosmology.
In other words we require the model in examination to address
questions like ``Is there a Dark Matter (DM) candidate? How abundant
is this at present?'' or  ``Can we find successful 
mechanisms for Baryogenesis and Reheating?''. These
questions arise from the more general program of using present
cosmological data to constrain the enormous proliferation of 
phenomenological works describing physics beyond the Standard Model
(SM). 

We are going to concentrate on one of such models, known as
``trinification'', introduced for the first time 
in 1984 by S. Glashow~\cite{Glash}, successively
studied in detail by Babu et al.~\cite{Babu} and more recently by
Willenbrock et al.~\cite{Wil}\cite{Will}. In particular we consider
the minimal formulation of the model (respect to how the SM is
embedded in it), and focus on the question of Dark Matter. 
We show that the model does {\em not} have a stable DM candidate
compatible with $\Omega_{DM} \leq 0.24$. We find that if we adjust the
parameters such that we have a stable candidate, there is far too much
DM. On the other hand, making the candidate unstable
conflicts with Big Bang Nucleosynthesis (BBN) constraints.
In section 2 we are going to
briefly present the model, list its salient features and focus on its
advantages and downsides. We present our results in section 3. 
We conclude in section 4 discussing possible improvements of the model
that would circumvent its difficulty in providing a viable DM candidate.

\section{Trinification in a nutshell}
The name ``trinification'' comes from its gauge symmetry: a triple
replica of $SU(3)$ conventionally written as $SU_C(3)\times SU_L(3)\times
SU_R(3)\times \mathbb{Z}_3$. The discrete group $\mathbb{Z}_3$ guarantees the gauge
couplings of the single $SU(3)$ factors are the same at the GUT scale. 
The SM embedding
is obtained by identifying the $SU_C(3)$ with the QCD gauge symmetry
while the electroweak gauge group emerges as a result of breaking the
other two $SU(3)$ factors. Each SM fermion generation is embedded in a
$\textbf{27}=(1,3,\bar{3})\oplus (\bar{3},1,3)\oplus (3,\bar{3},1)$
representation of the gauge group. In order to better
understand the field content we can re-express this representation in terms
of SM quantum numbers
\begin{eqnarray*}
E \equiv (1,2,-\frac{1}{2}) \qquad & N_1 \equiv (1,1,0) \qquad 
& E^c \equiv (1,2,\frac{1}{2})\\
L \equiv (1,2,-\frac{1}{2}) \qquad & N_2 \equiv (1,1,0) \qquad  
& e^c \equiv (1,1,1)\\
Q \equiv (3,2,\frac{1}{6}) \qquad & u^c \equiv (\bar{3},1,\frac{2}{3}) \qquad
& d^c \equiv (\bar{3},1,-\frac{1}{3})\\
B \equiv (3,1,-\frac{1}{3}) \qquad & B^c \equiv (\bar{3},1,-\frac{1}{3}). \qquad &
\end{eqnarray*}
We immediately notice that each generation includes the usual lepton
and quark doublets and singlets plus some additional fields:
we get two additional lepton doublets $E$ and $E^c$, two
neutral singlets $N_1$ and $N_2$ and two quark singlets $B$ and $B^c$.
We will see that the doublets under the breaking of $[SU(3)]^3$ will
acquire a heavy mass ($\sim O(M_{GUT})$) at tree level 
and become Dirac fermions. The singlets, instead, remain light and 
require a more refined adjustment involving radiative corrections.

In order to have unification of the SM couplings at the GUT scale we
also need two copies of scalars both in the same $\Phi^a(\textbf{27})\equiv
\Phi_{\ell}^a(1,3,\bar{3})\oplus \Phi_{q^c}^a(\bar{3},1,3)\oplus \Phi_q^a(3,\bar{3},1) \quad a=1,2 \quad$
representation the fermions are in. Let's concentrate only on the 
$\Phi_{\ell}^a$ and write its field content in terms of SM quantum
numbers 
\begin{eqnarray*}
\phi_{1\ell}^a \equiv (1,2,-\frac{1}{2}) \qquad & \phi_{2\ell}^a \equiv (1,2,\frac{1}{2}) \qquad 
& \phi_{3\ell}^a \equiv (1,2,-\frac{1}{2})\\
S^a_{1\ell} \equiv (1,1,0) \qquad & S_{2\ell}^a \equiv (1,1,1) \qquad  
& S^a_{3\ell} \equiv (1,1,0)
\end{eqnarray*}
The breaking $[SU(3)]^3 \to SU_C(3)\times SU_L(2)\times U_Y(1)$ is obtained giving
vevs to some of the singlets $S_{i\ell}^a$. The most general choice being ($S^a_{2\ell}$ 
are electrically charged so they cannot assume a vev)
\begin{equation}
<S_{3\ell}^1>=v_1 \qquad <S_{1\ell}^2>=v_2 \qquad <S_{3\ell}^2>=v_3,
\end{equation}
with $v$'s $\sim O(M_{GUT}=10^{14}\, GeV)$~\cite{Wil}. In the same fashion the
electroweak symmetry is broken giving vevs to the electrically neutral 
components of the doublets $\phi_i^a$ charged under the $SU_L(2)$. 
A very general choice is
\begin{eqnarray}
<(\phi_{1\ell}^2)^0>=n_1 \qquad & <(\phi_{2\ell}^2)^0>=n_2 \qquad &
<(\phi_{3\ell}^2)^0>=n_3 \nonumber \\
<(\phi_{1\ell}^1)^0>=u_1 \qquad & <(\phi_{2\ell}^1)^0>=u_2, \qquad &
\end{eqnarray}
with $u_i,\, n_i\sim O(M_{EW})$. The other scalar fields $\Phi_{q^c}^a$ and
$\Phi_q^a$ do not acquire any vev since they carry color charge and would
break $SU_C(3)$. They are generically very heavy due to radiative
corrections to their mass terms  and do not show up in the low energy
spectrum of the theory. From now on we will assume their masses to be
of the order of $M_{GUT}$.
\subsection{A simple case}
In a simplified version of the the model, we set $n_1=n_2=n_3=v_3\equiv0$ 
but keep all other vevs ($u_1,u_2,v_1,v_2$) non-zero. The qualitative results
will be exactly equal to the more general case and suffice to illustrate
our point. A linear combination of the $\phi_{i\ell}^a$ and four of the
$S_{i\ell}^a$ are eaten by twelve of the  gauge bosons\footnote{The gauge
  bosons are in the $\mathbf{24}= (1,1,8)\oplus (1,8,1)\oplus (8,1,1)$ 
  adjoint representation of $[SU(3)]^3$}
that become heavy
with masses proportional to the $v_i$. Fine-tuning the quartic couplings,
it is possible to obtain at most 5 light ($\sim M_{EW}$) Higgs doublets. At
the same time Yukawa terms give masses for the fermions at the
tree level. In general, such terms can be built pairing two fermion
doublets and a scalar singlet or one scalar doublet contracting indices 
with one fermion doublet plus one one fermion singlet. In the first
case we end up with a heavy (mass $\sim M_{GUT}$) Dirac fermion
meanwhile in the second case the fermions are light (mass $\sim
M_{EW}$). Limiting our analysis to one fermion generation we obtain at
tree level
\begin{eqnarray}
m_B \simeq \sqrt{g^2_1v^2_1+g^2_2v^2_2} & \qquad \qquad & m_E \simeq
\sqrt{h^2_1v^2_1+h^2_2v^2_2}\nonumber \\
m_u = g_1u_2 \qquad & m_d\simeq g_1u_1  \qquad & m_e\simeq h_1u_1 \\
m_\nu= h_1u_2  \qquad & m_{N_1}= h_1u_2  \qquad & m_{N_2}\simeq
\frac{h^2_1u_1u_2}{m_E}, \nonumber 
\end{eqnarray}
where $h_a$ and $g_a$ are couplings associated with the Yukawa terms
proportional to $\phi_{i\ell}^a$. This spectrum has some
positive qualitative features and a negative one. As we have already
mentioned $B$, $B^c$ and $E$, $E^c$ pair up to become very heavy
Dirac fermions. The up and down quarks as well as the electron 
get different light masses. Unfortunately $N_2$, $N_1$ and $\nu$ are
also light, with the additional inconvenience that the last two seem to
pair up to form a light Dirac fermion. This is highly undesirable and
can be fixed invoking radiative correction induced by cubic scalar
couplings~\cite{Wil}\cite{Will}. Calculating the one loop 
contribution to mass terms for light fermions and using a seesaw 
mechanism the spectrum for these light neutral fields become
\begin{equation}
m_{N_{1,2}}\sim g_q^2F \qquad m_\nu\sim \frac{h^2_1u^2_2}{g_q^2F}
\end{equation}  
where $F\leq M_{GUT}$ is a factor of pure one-loop origin. In the
presence of other two fermion generations there are additional $N_i$'s
and they can mix up and appear in a sort of hierarchy where the
lightest of all can be pushed as far down as $10^5\, GeV$. 
\subsection{Light and darkness of Trinification}
We end this section by briefly reviewing the positive and 
negative features of this GUT model.
Thanks to the two sets of scalar fields (six weak doublets~$\phi^a_{i\ell}$)
this model achieves unifications of the SM gauge couplings at a scale
around $10^{14}\, GeV$. The mechanism is similar to that of generic 
SUSY SU(5) GUT (although this unifies at $10^{16}GeV$): the six Higgs contribution
to the $\beta$-function is equivalent to the contribution of the two SUSY
Higgs doublets and their fermion superpartners. The only difference
resides in the fact that in Trinification there aren't gauge boson
superpartners that are ultimately responsible of moving the GUT scale
upwards.

Usually, in other non-SUSY unification models there are
mass degeneracies between quarks and leptons since they come in the
same representation of the gauge group. In Trinification this is avoided since quarks
and leptons are in different representations in which the
$\textbf{27}$ is decomposed. In conclusion SM masses for fermions
can be arbitrarily adjusted through Yukawa couplings to fit the 
experimental values. From this point of view Trinification is not
any more predictive than the SM itself.
Masses for the scalars are not  protected and receive one loop
quadratic corrections requiring fine tuning for at least one Higgs
light doublet in $\Phi_{\ell}^a$. Trinification does not provide any
mechanism to solve the so called hierarchy problem.

Unlike SU(5), in trinification gauge bosons conserve Baryon 
number and so do not mediate
proton decay: the proton can only decay through Yukawa
interactions. An acceptable value for its lifetime is recovered
without the need of fine-tuning the involved Yukawa couplings.

Due to abundant number of scalar fields, it is possible that 
baryogenesis is achieved at GUT scales~\cite{Wei} or at electroweak 
scales through a first order phase transition~\cite{Cli}. Heavy 
scalars may be important in some Inflationary scenario.
Moreover light neutral singlets may play the role of sterile 
neutrinos and be used to invoke a Leptogenesis
mechanism~\cite{Fuku}. The punch line being that this model
offers a wide variety of possible developments in the context of
Cosmology. In the next section we will address one of 
these issues. We investigate the possibility that the lightest of the
neutral singlets can function as a viable candidate for Dark Matter 
and try to give an indicative answer.
\section{A Dark Matter Candidate}
Let us indicate the lightest neutral singlet as $N_\chi$. There are two 
main requirements that $N_\chi$ needs to satisfy in order to be a possible
Dark Matter candidate. First, its relative abundance $\Omega_{N_\chi}$ has to 
be less than or equal to 0.24. Second, its decay time needs to be much longer
than the age of the Universe.
\subsection{Relative abundance}
The relative abundance at present (assuming the particle is stable) 
is calculated using the Boltzmann equation. The form of its solution
depends on the regime (relativistic or not) at which it is
approximated. To check if $N_\chi$ was non-relativistic at
freeze out we need to verify under which condition the 
inequality $x_f=\frac{m}{T}\geq 3$ is valid. If we assume that $N_\chi$ is
non-relativistic the value of $x_f$ depends 
logarithmically on the annihilation cross section and the mass of
$N_\chi$~\cite{Kolb}.
\begin{equation}
\label{eq:crossec}
x_f \simeq \ln \left( 0.038 \frac{g}{\sqrt{g_*}}M_{PL}m_{N_\chi}\left< \sigma v\right> \right)
\end{equation}
Since we are interested only in estimating this expression 
we consider the dominant contribution to the annihilation channel. 
This is given by the following tree level Feynman graph\footnote{The
  contribution coming from two $N_\chi$ singlets annihilating in a
  virtual heavy $X$ boson is even more suppressed since it is 
  proportional to $\frac{\alpha^2_{gauge}}{m^2_X}$ and $m_X^2 \sim v^2_ig_{gauge}$.}

\begin{equation}
\psset{unit=1mm,linecolor=black}
\def\freccina{%
             \psline[linewidth=.1](-1.5,1.5)(0,0)(-1.5,-1.5)
             }
\begin{pspicture}[](-30,-17)(50,17)
\psline[linewidth=.2](-29,15)(10,15)(10,-15)(-29,-15)
\psline[linewidth=.2,linestyle=dashed](10,15)(49,15)\relax
\psline[linewidth=.2,linestyle=dashed](10,-15)(49,-15)\relax
\rput{0}(-14,15){\freccina}
\rput{270}(10,0){\freccina}
\rput{180}(-14,-15){\freccina}
\rput{0}(-25,12){$N_\chi$}
\rput{0}(-25,-12){$\bar{N_\chi}$}
\rput{0}(6,0){$E$}
\rput{0}(46,12){$\phi_{i\ell}^a$}
\rput{0}(46,-12){$\phi_{i\ell}^{\ast a}$}
\end{pspicture}.
\end{equation}

Assuming $\phi_{i\ell}^a$ and $\phi_{i\ell}^{\ast a}$ are light Higgs we are lead to 
the s-wave expression for the cross-section 
\begin{equation}
\left< \sigma v\right> \sim \frac{h^4}{(4\pi)^2m^2_E}.
\end{equation}
Here $h$ is a generic Yukawa coupling. Plugging this expression in
(\ref{eq:crossec}) and assuming the reasonable range $10^{5}\, GeV \leq
m_{N_\chi}\leq 10^{10}\, GeV$ we arrive at the conclusion that
$\frac{v_i}{h}\leq 10^9\, GeV$ in order to have $x_f\geq 3$. Since $v_i \sim
M_{GUT}$ the inequality cannot be satisfied implying 
that $N_\chi$ is highly relativistic when it freezes out.
 
We then equate the annihilation rate of $N_{\chi}$ 
to the Hubble rate $H$ to obtain the freeze-out temperature
\begin{equation}
T_f \approx 9.1 \frac{\sqrt{g_{*}(T_f)}}{M_{Pl}\langle\sigma v\rangle}\sim 10^{12}\sqrt{g_{*}}h^{-2}\, GeV.
\end{equation}
This gives a present-day abundance that is rather insensitive to 
$\langle\sigma v\rangle$ and much too large, namely
\begin{equation}
 \Omega_{N_\chi} \sim 2\times 10^{8}g_{*S}^{-1}(T_f)\frac{m_{N_\chi}}{GeV}.
\end{equation} 
\subsection{Decay time}
The previous estimate is based on the assumption that $N_{\chi}$ is
stable over the course of the universe requiring that
$\tau_{N\chi} \gg  10^{10}\, yr$. 
However, this is not correct for most reasonable choices of 
parameters in this model. We find that it decays not too long 
after it freezes out, and tends to destroy Big Bang Nucleosynthesis 
(BBN) unless we adjust some parameters appropriately. In order 
to estimate the lifetime of $N_{\chi}$ we will consider only the 
dominant contribution, as we did earlier for the cross section. 
The most favorable decay channel is given by the following

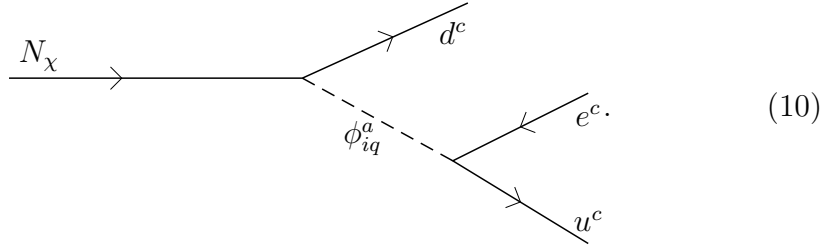
\begin{equation}
\psset{unit=1mm,linecolor=black}
\def\freccina{%
             \psline[linewidth=.1](-1.5,1.5)(0,0)(-1.5,-1.5)
             }
\begin{pspicture}[](-30,-17)(50,17)
\psline[linewidth=.2](-29,5)(10,5)(32,15)
\psline[linewidth=.2,linestyle=dashed](10,5)(30,-6)\relax
\psline[linewidth=.2](48,3)(30,-6)(48,-17)\relax
\rput{0}(-14,5){\freccina}
\rput{24}(22,10.5){\freccina}
\rput{204}(39,-1.5){\freccina}
\rput{336}(39,-11.5){\freccina}
\rput{0}(-25,8){$N_\chi$}
\rput{0}(30,11){$d^c$}
\rput{0}(18,-3){$\phi_{iq}^a$}
\rput{0}(48,0){$e^c$}
\rput{0}(48,-14){$u^c$}
\end{pspicture}.
\end{equation}

In this graph $\phi_{iq}^a$ is one of the scalars carrying color charge
with mass of the order of $M_{GUT}$. In calculating the decay rate it
is a very good approximation to consider the decay products massless
since they are highly relativistic. In such an approximation the decay
rate is
\begin{equation}
\Gamma_{N_\chi} \sim \frac{128g^4_q}{(2\pi)^2} \frac{m^5_{N_\chi}}{m_{\phi_{iq}^a}^4}
\end{equation}
with $g_q$ being the Yukawa coupling associated with the heavy colored
scalar. For $m_{\phi_{iq}^a} \sim 10^{14}\, GeV$ and $m_{N_\chi} \sim 10^5\, GeV$
we get the following estimate for the decay time
\begin{equation}
\tau_{N_\chi} \sim 10^{30}g^{-4}_q\, GeV^{-1} \sim 10^6g^{-4}_q\, s.
\end{equation}
For a typical value of $g_q=0.1$ the decay time is about a 600 years making
$N_\chi$ rather unstable and completely absent at present time in the
universe. Moreover its decay products are so highly relativistic and
get produced in such abundance that they destroy any product of
BBN. We can reverse the reasoning and put an upper bound on the Yukawa
coupling that stabilizes $N_\chi$
\begin{equation}
\tau_{N_\chi}= 6.3 \times 10^{-2}g^{-4}_q\, yr \gg  10^{10}\, yr \qquad \Longrightarrow \qquad
g_q \ll  2 \times 10^{-3}.
\end{equation}
This fine-tuning of $g_{q}$ is not so unreasonable, but it may backfire on the
one-loop corrections to the mass of the neutral singlets compromising
the efficiency of the radiative seesaw mechanism.    
\section{Conclusion and discussion}
We have seen that the lightest of the neutral singlets in the
minimal trinification model is greatly overabundant at freeze out with
the consequence that it is completely ruled out by
cosmological data. This may be
overcome if the particle is not very stable but then it ruins BBN and
creates another big problem. As it stands this GUT model does not seem
to be a complete model of particle physics since it does not withstand
one of the most needed cosmological requirements: the
existence of Dark Matter. 

There is room for improvement though. On one
side we may introduce a $\mathbb{Z}_2$ symmetry like the one
introduced in SUSY models and make $N_\chi$ more stable without the need
of adjusting the Yukawas opportunely. Such a discrete symmetry would
forbid the Yukawa terms responsible for the dominant decay channel and
stabilize the singlet. On the other hand we have to admit that our
calculation of the cross-section for the relative abundance is a
little ``naive'' since it doesn't take into account possible mixing of
$N_i$'s with the other neutral Weyl fermions of the model. For example
within just one generations of leptons we have $E_0 \in E$ and $E_0^c
\in E^c$ plus the two neutral singlets, $N_1$ and $N_2$, and the
neutrino $\nu$. The mass 
matrix of these five neutral fields is far from diagonal and can 
provide some mixing between them. The net result is that the lightest
among them may have some non-zero coupling with the $Z^0$ boson
greatly enhancing the cross section amplitude. This mechanism may help
in reducing the abundance to more accepted values.

Finally we would like to thank the organizers of the LXXXVI 
``Les Houches'' summer school, the staff and all our colleagues at the
school that made our stay enjoyable and this work possible.

\end{document}